\documentclass{icrc2009}

\usepackage{graphicx}   
\usepackage[caption=false]{caption}    
\usepackage[font=footnotesize]{subfig} 
\usepackage{fixltx2e}
\usepackage{url}

\newcommand{\shorttitle}[1]%
{\markboth{Proceedings of the 31\MakeLowercase{$^{st}$} ICRC, {\L}\'{o}d\'{z} 2009}{#1} }
\newcommand{\etal}{\MakeLowercase{\textit{et al. }}} 

\begin{document}
\title{VERITAS Observations of Magnetars}

\author{\IEEEauthorblockN{Roxanne Guenette\IEEEauthorrefmark{1} for the VERITAS collaboration\IEEEauthorrefmark{2}}
\\
\IEEEauthorblockA{\IEEEauthorrefmark{1}Department of Physics, McGill University, H3A 2T8 Montreal, Qc, Canada (guenette@physics.mcgill.ca)}
\IEEEauthorblockA{\IEEEauthorrefmark{2}see R.A. Ong et al. (these proceedings) or http://veritas.sao.arizona.edu/conferences/authors?icrc2009}}

\shorttitle{R.Guenette \etal VERITAS Observations of Magnetars}
\maketitle

\begin{abstract}

 Magnetars are rotating neutron stars with extremely strong magnetic fields ($\sim 10^{14}-10^{15}$G). X-ray and soft gamma-ray observations have revealed the existence of non-thermal particle populations which may suggest emission of very high energy photons. VERITAS, an array of four 12m imaging atmospheric Cherenkov telescopes, is designed to observe gamma-ray emission between 100 GeV and 30 TeV. Here we present the results of VERITAS observations of two magnetars, 4U 0142+61 and 1E 2259+586.\\

\end{abstract}

\begin{IEEEkeywords}
Gamma rays, observations, magnetars 
\end{IEEEkeywords}
 
\section{Introduction}

Magnetars are a subclass of neutron stars, including anomalous X-ray pulsars (AXPs) and soft gamma repeaters (SGRs), characterised by ultra-high magnetic fields ($\sim 10^{14}-10^{15}$G). There are about 15 known magnetars \cite{WoodsThompsoncatalogue}. All share these main characteristics: persistent X-ray luminosity of the order of $10^{34}-10^{36}$erg/s, slow X-ray pulsations between 2s and 12s and large spin-down rates between $10^{-10}$ and $10^{-12}$ss$^{-1}$. Their persistent X-ray luminosity exceeds the spin-down luminosity, implying that the X-ray emission cannot be explained by the rotational energy losses alone. The magnetar model has been proposed \cite{DuncanandThompson1992,ThompsonandDuncan1995,ThompsonandDuncan1996} to explain the behaviour of SGRs, which undergo occasional outbursts. It is now clear that SGRs and AXPs share the same properties and it is widely accepted that they both can be classified as magnetars: neutron stars powered by magnetic-field decay.\\

The X-ray spectra from different magnetars have unambiguously shown a non-thermal component \cite{denHartog2006,Woods2004}. Different theoretical models have been proposed to explain the presence of the non-thermal emission \cite{Rea2008,Fernandez2007,Thompson2005}. These models suggest mechanisms that purport to reproduce the spectra in X-ray and soft gamma-ray energy bands; none of these models make predictions for steady emission in the TeV range. \\

Magnetars exhibit bursting activity in the X-ray (small and giant flares) which can be explained by fractures of their crust from magnetic field rearrangements \cite{ThompsonandDuncan1995}. It has been proposed that during SGR giant outbursts, TeV gamma rays could be produced in electromagnetic processes and also in hadronic interactions with the thermal radiation \cite{Halzen2005}. Halzen et al. calculated the TeV gamma-ray flux produced in a giant burst for a spectral index of -2.00 and obtained 8.23 x $10^{-7}$cm$^{-2}$s$^{-1}$ for the 0.3s duration of the burst, corresponding to $\sim$100 times the Crab Nebula flux. Target of Opportunity observations with air Cherenkov telescopes, triggered by bright X-ray flares, can be used to constrain such theoretical conjecture.\\ 

Two magnetars are presented here: 4U 0142+61 and 1E 2259+586. 4U 0142+61 is an isolated magnetar at a distance of 3.6 kpc \cite{Durant2006}. The magnetar's period is 8.69s and the period derivative is 0.2 x $10^{-11}$ss$^{-1}$ \cite{4U_highenergy}\cite{guver}. Its magnetic field is 1.3 x 10$^{14}$G. This magnetar entered a state of X-ray bursting activity in April 2006 which lasted until February 2007 \cite{Gavriil_4U}. During this active state, several bursts were detected (one in April 2006, four in June 2006 and one in February 2007). Furthermore, this magnetar might be surrounded by a gaseous fallback disk \cite{ertan_4U}. VERITAS observed this source in October 2008. \\

1E 2259+586 is a magnetar surrounded by a large (0.5$^{\circ}$) supernova remnant (CTB 109) at a distance of approximately 3 kpc \cite{Kothes2002}.  The magnetar's period is 6.98s, its period derivative is 4.84 x 10$^{-13}$ss$^{-1}$ and its magnetic field is 6.3 x 10$^{13}$G. The magnetar was observed during a major X-ray outburst in June 2002 which lasted for 4h \cite{Kaspi2003}\cite{Woods2004}. During this burst, significant changes from the persistent emission were observed in the spectra and in the timing. The characteristics of these two objects are summarised in Table \ref{table_char}.\\

\begin{table}[th]
  \caption{Characteristics of the magnetars presented.}
  \label{table_char}
  \centering
  \begin{tabular}{|c|c|c|}
   \hline
   Object                                     &  4U 0142+61    &   1E 2259+586 \\
   \hline
   Magnetic field (G)                         & 1.3x$10^{14}$  &   6.3x$10^{13}$ \\
   \hline
   Distance (kpc)                             &     3.6        &   $\sim$ 3       \\
   \hline
   Period (s)                                 &    8.69        &   6.98         \\
   \hline
   Period derivative  ( $10^{-13}$ ss$^{-1}$) &     20         &   4.84         \\
   \hline
   \end{tabular}
  \end{table} 

\section{Observations}
 
VERITAS, the Very Energetic Radiation Imaging Telescope Array System, is an array of four 12m imaging atmospheric Cherenkov telescopes located in Southern Arizona, USA. It is designed to observe VHE gamma-ray emission between 100 GeV and 30 TeV. The energy resolution is from 15$\%$ to 20$\%$ and the angular resolution is $\sim$ 0.1$^{\circ}$. The VERITAS camera field of view is 3.5$^{\circ}$. The high sensitivity of the experiment allows the detection of sources with a flux of 1$\%$ of the Crab Nebula in less than 50 hours of observations. For more details see \cite{Acciari2008}.\\

VERITAS observations of 4U 0142+61 were performed in October 2007 during low to moderate moonlight and during dedicated dark time observations of a triggered gamma-ray burst, which was between 0.4$^{\circ}$ and 1.1$^{\circ}$ from 4U 0142+61. The magnetar was in quiescence during these observations. The observations of 1E 2259+586 were taken at the end of December 2007 and at the beginning of January 2008, during dark time dedicated to the supernova remnant (SNR) surrounding it (CTB 109), in \textit{wobble} mode \cite{fomin94}, in which the source is 0.5$^{\circ}$ away from the camera centre in different directions. No X-ray burst was observed at the moment of the observations.

\section{Analysis and Results}

For this analysis, only runs passing quality cuts (good weather and no hardware problems) were selected. The data were taken with three or four telescopes. The analysis was performed using the standard second-moment parametrisation of the telescope images \cite{Hillas1985}. Requirements for the images are: a minimum of 500 digital counts ($\approx$ 94 photoelectrons) in the image, less than 10$\%$ of the image in the pixels at the edge of the camera and the image centre of gravity less than 1.5$^{\circ}$ from the camera centre. Following cuts were applied: -1.2 $<$ reduced mean scale width/length $<$ 0.5 and $\theta^{2}$ $<$ 0.015 deg$^{2}$. The background region was estimated using the ``ring background'' model \cite{Aharonian2005}.\\

The significance calculations were done using equation 17 from Li and Ma \cite{LiandMa}. Results are listed in Table \ref{table_results}, where N$_{on}$ is the number of events at the source position, N$_{off}$ is the estimated number of background events at the source position and $\alpha$ is the normalisation factor between N$_{on}$ and N$_{off}$. The 99$\%$ confidence level (C.L.) flux upper limits for energies above 400 GeV were calculated using the Helene method \cite{Helene}, assuming a power law energy spectrum of index -2.5, and are also expressed relative to the Crab Nebula flux. The results from the analysis presented here are summarised in Table \ref{table_results}.\\

  \begin{table*}[th]
  \caption{Results on the magnetar observations.}
  \label{table_results}
  \centering
  \begin{tabular}{|c|c|c|c|c|c|c|c|}      
  \hline
   Object      & Exposure & N$_{on}$ & N$_{off}$  & $\alpha$ & Significance  & 99$\%$ C.L. Upper limit(E$>$400 GeV) \\
               &   (h)    &          &            &          &   ($\sigma$)  &   (cm$^{-2}$ s$^{-1}$)   [Crab Units]   \\
   \hline                                                                                                         
   4U 0142+61  &    9.25  &     82   &   454      &   0.25   &     -2.9      &            8.68x$10^{-13}$ [0.9$\%$]   \\
   1E 2259+586 &    4.27  &     40   &   185      &   0.23   &      1.5      &            2.49x$10^{-12}$ [2.5$\%$]   \\
  \hline
  \end{tabular}
  \end{table*}

\noindent
\textit{4U 0142+61}

\begin{figure}[h!t]
  \centering
  \includegraphics[width=3.0in]{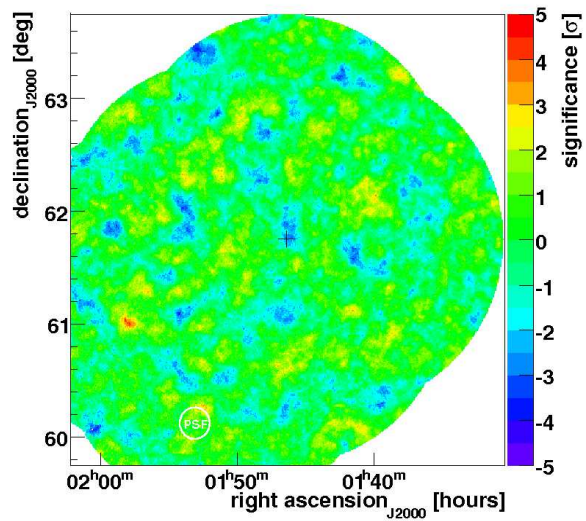}
  \caption{Significance skymap for 4U 0142+61. The black cross shows the source position and the white circle at the bottom left represents the point spread function (psf) of the instrument.}
  \label{fig_4U}
 \end{figure}

No gamma-ray emission has been detected by VERITAS. The significance skymap is shown in Figure \ref{fig_4U}. The analysis gives a 99$\%$ C.L. flux upper limit less than 0.9$\%$ of the Crab Nebula. This result is for the quiescent phase, since the magnetar has not been detected in an active state since 2007.\\

\noindent
\textit{1E 2259+586}

\begin{figure}[h!t]
  \centering
  \includegraphics[width=3.0in]{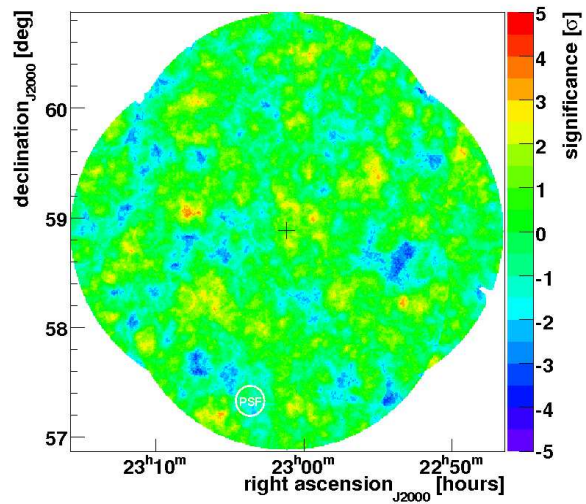}
  \caption{Significance skymap for 1E 2259+586. The black cross shows the source position and the white circle at the bottom left represents the psf of the instrument.}
  \label{fig_1E2259}
 \end{figure}

The VERITAS observations show no gamma-ray emission from 1E 2259+586. The significance skymap is shown in Figure \ref{fig_1E2259}. A 99$\%$ C.L. flux upper limit of less than 2.5$\%$ of the Crab Nebula has been estimated for the quiescent phase. Since this magnetar is at the centre of the SNR CTB 109, the result presented here can also be applied to point source emission from the SNR.\\

\section{Discussion and Conclusion }

No gamma-ray emission was detected by VERITAS from either of the
magnetars presented here. Steady VHE gamma-ray emission is not
predicted by any theoretical models on magnetars, and such detection
would have been unexpected. \\

TeV emission during a major X-ray outburst has been proposed by
\cite{Halzen2005} in which the VHE gamma-rays are produced in
electromagnetic processes or in hadronic interactions with the thermal
radiation. The flux predicted during a short burst (0.3s) is the order of 100
times the Crab Nebula flux. Such a flux, if observed during the
outburst would be easily detectable by air Cherenkov telescopes.\\

Observation of magnetars in the VHE regime during a major outburst
would give useful constraints to the non-thermal emission during such
bursts, as suggested in \cite{Halzen2005}. Target of Opportunity
programs at VERITAS include several magnetars.\\

\section{Acknowledgements}

This research is supported by grants from the US Department of Energy,
the US National Science Foundation, and the Smithsonian Institution,
by FQRNT and NSERC in Canada, by Science Foundation Ireland, and by
STFC in the UK. We acknowledge the excellent work of the technical
support staff at the FLWO and the collaborating institutions in the
construction and operation of the instrument.\\


\end{document}